# Visual Interface Workflow Management System Strengthening Data Integrity and Project Tracking in Complex Processes

**Ömer Elri[1*]**　　　　　　　　　　　　　　**Doç. Dr. Serkan Savaş[1]**

[1] *Kırıkkale University, Engineering and Natural Sciences Faculty, Computer Engineering Department, Kırıkkale, Türkiye*

**\*E-mail: omer@omerelri.com**　　　　　　　　　　　　**\*Orcid No: 0009-0003-8032-109X**

**Abstract**

Manual notes and scattered messaging applications used in managing business processes compromise data integrity and abstract project tracking. In this study, an integrated system that works simultaneously on web and mobile platforms has been developed to enable individual users and teams to manage their workflows with concrete data. The system architecture integrates MongoDB, which stores data in JSON format, Node.js/Express.js on the server side, React.js on the web interface, and React Native technologies on the mobile side. The system interface is designed around visual dashboards that track the status of tasks (To Do/In Progress/Done). The urgency of tasks is distinguished by color-coded labels, and dynamic graphics (Dashboard) have been created for managers to monitor team performance. The usability of the system was tested with a heterogeneous group of 10 people consisting of engineers, engineering students, public employees, branch managers, and healthcare personnel. In analyses conducted using a 5-point Likert scale, the organizational efficiency provided by the system compared to traditional methods was rated 4.90, while the visual dashboards achieved a perfect score of 5.00 with zero variance ($\sigma=0.00$). Additionally, the ease of interface use was rated 4.65, and overall user satisfaction was calculated as 4.60. The findings show that the developed system simplifies complex work processes and provides a traceable digital working environment for Small and Medium-sized Enterprises and project teams.

**Acknowledgements:** This work is supported by the Scientific and Technological Research Council of Türkiye (TÜBİTAK) under the 2209-A University Student Research Projects Support Program (Project No: 1919B012453439, 2024/1 term). The authors are solely responsible for the content of this study, and the views expressed do not necessarily reflect the official positions of TÜBİTAK.

## 1. INTRODUCTION

With the rapid advancement of technology and the widespread adoption of digitalization processes today, the complexity of project management has increased exponentially. Efforts to digitize project management processes are developing, particularly with examples focused on the software and defense industries. In software projects, the integration of risk management, decision support, and process monitoring functions with web-based platforms and artificial intelligence stands out as one of the early and typical application areas of digital project management. In the defense industry, corporate content and document management infrastructures are used to integrate documents produced throughout the project life cycle into a single digital platform and to increase the traceability of processes. These efforts demonstrate that digitalization is not only a technical automation but also a transformation tool that standardizes information flow among project stakeholders and strengthens corporate memory [1].

Studies focusing on software projects aim to transfer project risk management to the digital environment and support it with artificial intelligence techniques such as artificial neural networks, enabling the early prediction of deviations in project outputs and the taking of preventive decisions. Such systems enrich project management processes with digital dashboards, automated alerts, and data-driven decision support mechanisms, offering a dynamic and learning structure that goes beyond traditional risk lists [2].

Workflow Management Systems (WMS) are positioned as software solutions that enable the definition, modeling, execution, and monitoring of corporate processes, often integrated with electronic document and content management infrastructures. Studies conducted in Türkiye, particularly research carried out on Electronic Document Management Systems (EDMS), reveal that workflow engines standardize corporate operations by automating role-based routing, time stamping, record integrity, and authorization functions in processes such as document production, approval, transfer, archiving, and access. In this context, WMS is considered an infrastructure element that supports the document lifecycle within the framework of the TS 13298 standard, transfers process steps to the digital environment, and facilitates timely managerial decision-making [3, 4].

Engineering teams and software developers face the challenge of managing multiple tasks, deadlines, and team interactions simultaneously. Traditional project management methods, such as physical notebooks, whiteboard systems, or disconnected spreadsheet software, often lead to "data silos." These methods lack real-time synchronization. This makes it difficult to monitor the status of the project, leading to data loss, communication breakdowns, and decreased efficiency.

The primary objective of this study is to develop a comprehensive WMS that addresses these limitations. The proposed system aims to digitize the entire project lifecycle and ensure data integrity and accessibility across different platforms. Developed using modern web technologies and mobile accessibility, the system allows team leaders to visually assign tasks, monitor progress through intuitive dashboards, and receive real-time updates.

The remainder of the study is structured as follows. The second section describes the materials and methods used, while the third section presents and discusses the results. The final section concludes the study and makes recommendations.

## 2. MATERIALS AND METHODS

### 2.1. System Architecture and Technologies

The system is built on a client-server architecture to ensure scalability and performance. The MERN Stack, which provides a unified JavaScript development environment, has been chosen as the core technology stack for development. The MERN Stack is a popular JavaScript-based full-stack development technology stack used to build modern web applications. It consists of the MongoDB database, Express.js and Node.js server-side frameworks, and the React.js user interface library. This allows developers to use JavaScript, a single language, for both front-end and back-end development [5].

- **Database (MongoDB):** A NoSQL database was selected due to its flexibility in processing unstructured data such as task descriptions, user comments, and activity logs. It also offers scalability [6].

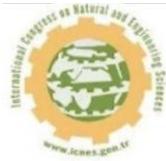

- **Back End (Node.js & Express.js):** The server-side application has been developed using Node.js, which is suitable for processing concurrent user requests. Node.js is an event-driven, single-threaded, and asynchronous JavaScript runtime environment. It provides high performance on the server side for scalable network applications. Express.js is a minimalist web framework built on Node.js. It simplifies functions such as routing, middleware, and HyperText Transfer Protocol (HTTP) request management, enabling rapid Application Programming Interface (API) development. Frequently used in academic studies within the MERN stack, this duo is preferred for real-time data processing in educational platforms and collaborative tools [7-9]. Node.js and Express.js-based frameworks demonstrate superior performance with low latency and high throughput compared to alternatives such as Hapi and Koa. Research confirms that these technologies provide gains of up to 74% in horizontal scalability [10]. RESTful API endpoints have been created to manage the data flow between the client and the database. Authentication processes are encrypted with JSON Web Tokens (JWT) to ensure data security.

- **Web Front End (React.js):** The desktop interface was built using React.js with a component-based architecture. React.js is an open-source JavaScript library developed by Facebook and employs a component-based, declarative approach for creating user interfaces. Thanks to Virtual Document Object Model (DOM) technology, it minimizes real DOM manipulation and provides high-performance rendering. This feature offers advantages, especially in developing dynamic Single Page Applications (SPA) [11]. Global state management has been implemented with Redux Toolkit to ensure that task updates are immediately reflected in the user interface without refreshing the page.

- **Mobile Application (React Native):** A cross-platform mobile application has been developed using React Native to support remote work. React Native is an open-source mobile application development framework based on React.js that enables the creation of native-performance applications on both iOS and Android platforms using a single JavaScript code base. Thanks to the bridge mechanism that enables JavaScript code to interact with native code such as Objective-C/Swift (iOS) and Java/Kotlin (Android), it accelerates cross-platform development processes and offers reusability [12]. This allows users to access the system from both Android and iOS devices and stay connected to their workflow while on the move.

## 2.2. Development Process

The project was carried out using an iterative process consisting of analysis, design, coding, and testing phases, in accordance with Software Development Life Cycle (SDLC) standards. In the first phase, requirements analysis was performed to determine the system's basic functions, followed by the creation of the database schema and user interface designs. During the development process, the backend services were prepared first, and then the web and mobile interfaces were integrated with these services. In the final stage, the system was tested with real data and optimized.

## 2.3. Test Process

In this study, quantitative research methods were used to analyze the user experience (UX) and technical adequacy of the developed application. The system's usability was tested with a heterogeneous group of 10 people with different demographic and professional backgrounds, consisting of engineers, engineering students, public employees, branch managers, and healthcare personnel. This participant profile ensured a comprehensive assessment of how the application performed across different levels of digital literacy and various workflows.

During the data collection process, participants completed a structured Usability Questionnaire immediately after their application experience. The questionnaire was designed to cover key dimensions such as the learnability of the interface, ease of navigation, system performance, data synchronization, and the effectiveness of functional features (task management, dashboard, team management). Ratings were collected using a 5-point Likert Scale (1: Strongly Disagree, 5: Strongly Agree), and an open-ended question was added to gather participants' qualitative opinions. Table 1 shows the survey items and which usability dimension in the literature each item focused on.

Table 1: Usability and Functional Evaluation Scale

| Item No | Evaluation Statement (Survey Question) | Dimension / Focus Area |
|---|---|---|
| Q1 | The application interface (Web and Mobile) is clear and user-friendly. | User Interface (UI) |
| Q2 | I can easily access the menus (Tasks, Dashboard, Team, etc.) within the application. | Navigation & Accessibility |
| Q3 | The operating speed and response time of the application are sufficient. | System Performance |
| Q4 | Data synchronization between the web interface and the mobile app is fast and reliable. | Data Synchronization |
| Q5 | The process of creating and detailing a new task is practical. | Operational Efficiency |
| Q6 | It is easy to update the status of tasks (To-Do, In Progress, Completed). | Task Management |
| Q7 | The ability to add visuals/files (Assets) to tasks facilitates workflow tracking. | Functional Utility |
| Q8 | The "Trash" feature provides confidence in managing accidentally deleted tasks. | Error Management & Trust |
| Q9 | The charts and summary data on the Dashboard made it easier to understand the project status. | Data Visualization |
| Q10 | Prioritizing tasks (High, Medium, Low) helped me organize my workflow. | Decision Support |
| Q11 | The app makes task tracking more organized compared to traditional methods (paper, WhatsApp, etc.). | Comparative Advantage |
| Q12 | The panel for managing team members and roles (Admin/User) is functional. | Administrative Functionality |
| Q13 | I would consider using this application for my own projects or business tracking. | Intent to Use |
| Q14 | Are there any features you like most or any parts you suggest for improvement? | Qualitative Feedback |

## 3. RESULTS AND DISCUSSION

The system developed within the scope of this study has been developed as web and mobile-based to provide sustainable project management without time and space constraints. The system's strongest feature is the preservation of interface consistency on both platforms and the seamless continuation of UX.

### 3.1. Web Interface and Central Management

The web interface is designed as a command center that allows project managers and team members to monitor the overall status of the project from a single screen. The designed control panel is presented in Figure 1.

Figure 1: Dashboard summarizing project statistics and real-time task statuses

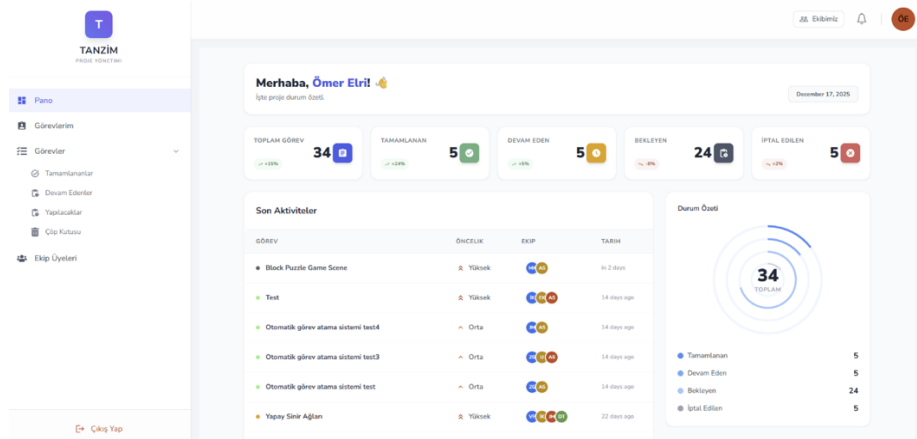

The main control panel shown in Figure 1 displays the total workload of the project, the number of completed tasks, and active processes in the form of summary cards. This structure enables managers to quickly analyze the status of the project with numerical data and manage the distribution of work in a balanced manner. Task assignments, prioritizations, and date updates made through the web interface are processed into the database and updated across the entire ecosystem. Users can create tasks using a simple and intuitive interface without having to deal with complex forms.

### 3.2. Mobile Application Findings for High Accessibility and Mobile Integration

The mobile application aims to ensure business processes continue without interruption by working in full synchronization with the data on the web platform. The application design has been developed adhering to the UX principles of the web interface, ensuring that users do not experience any adaptation issues when switching between different devices. The mobile part of the system prioritizes continuity in project management without being tied to the office environment. Figure 2 shows sample images from the system's mobile interface.

**Figure 2: Mobile Application Screens**

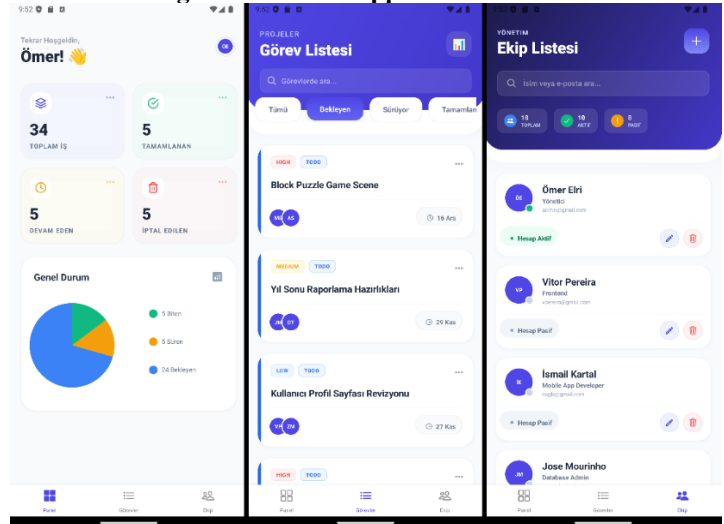

The interfaces shown in Figure 2 demonstrate the application's ergonomic and user-friendly design. Figure 2 shows the home page, real-time task list tracking, and profile/team management screens. This structure allows users to update their task statuses in real time and access project data at any time. The responses to the survey questions administered to users are presented in Table 2.

**Table 2: Quantitative Analysis of User Evaluation (N=10)**

| Item | Evaluation Metric | Mean ($M$) | Std. Dev. ($\sigma$) |
|---|---|---|---|
| Q1 | User Interface (UI) | 4.60 | 0.52 |
| Q2 | Navigation & Accessibility | 4.70 | 0.48 |
| Q3 | System Performance | 4.30 | 0.95 |
| Q4 | Data Synchronization | 4.00 | 0.94 |
| Q5 | Task Creation Process | 4.60 | 0.70 |
| Q6 | Status Updates | 4.60 | 0.97 |
| Q7 | Asset (File/Image) Addition | 4.30 | 1.16 |
| Q8 | Error Management (Trash Feature) | 4.80 | 0.42 |
| Q9 | Dashboard & Data Visualization | 5.00 | 0.00 |
| Q10 | Task Prioritization | 4.60 | 0.70 |
| Q11 | Advantage over Traditional Methods | 4.90 | 0.32 |
| Q12 | Team & Role Management | 4.90 | 0.32 |
| Q13 | Intent to Use / Adoption | 4.60 | 0.70 |

The usability testing yielded highly positive results across all dimensions. The system achieved a perfect score in Dashboard and Data Visualization ($M = 5.00$, $\sigma = 0.00$), indicating that the graphical summaries effectively provide project oversight. Furthermore, participants strongly agreed that the application provides a superior alternative to traditional methods such as paper-based tracking or instant messaging groups ($M = 4.90, \sigma = 0.32$).

While the overall performance was rated high, Data Synchronization ($M = 4.00$) and Asset Addition ($M = 4.30$) showed slightly higher standard deviations ($\sigma = 0.94$ and $\sigma = 1.16$, respectively). This suggests that while most users found these features functional, a small subset of the heterogeneous group (likely due to different mobile devices or connection speeds) experienced minor variances in performance. The results of user feedback are presented in Figure 3.

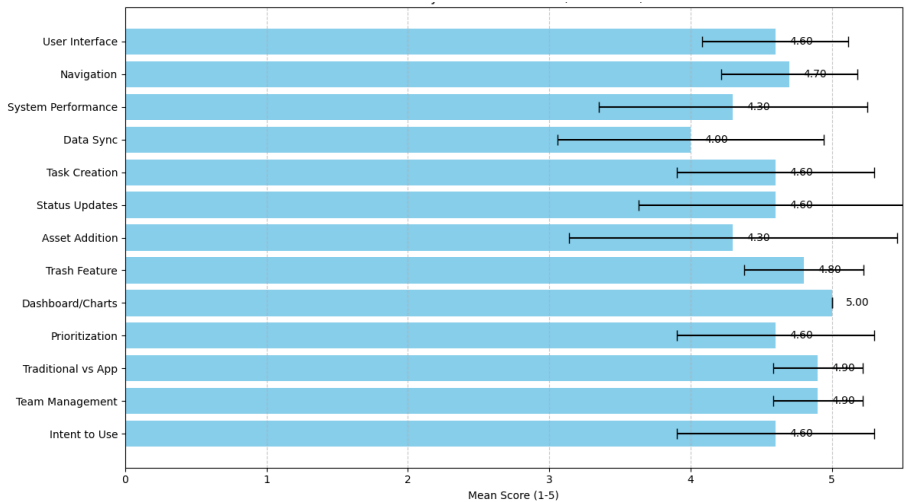

**Figure 3: Usability Evaluation Results (Mean $\mp \sigma$)**

Qualitative responses highlighted the "simplicity of the interface" and "practicality for corporate use." Some users suggested improvements in "responsive design for pop-up screens" and "increasing file upload capacities," which are noted for future development iterations. Some users, however, cited the seamless transition between web and mobile and the convenience of being able to "check on work I started in the office while on the go" as the system's greatest advantage. The results in Table 3 are a summary of the responses given to the survey.

**Table 3: User Feedback Analysis (N=10)**

| Evaluation Criterion | Mean ($M$) | SD ($\sigma$) | Qualitative Interpretation |
|---|---|---|---|
| Interface Ease of Use | 4.65 | 0.50 | Highly User-Friendly |
| Efficiency vs. Traditional Methods | 4.90 | 0.32 | Significantly More Organized |
| Utility for Task Management | 4.60 | 0.70 | Highly Beneficial |
| Error Mitigation & Reliability | 4.80 | 0.42 | Effective in Risk Reduction |
| Dashboard & Data Visibility | 5.00 | 0.00 | Excellent Visibility |
| Overall Satisfaction | 4.60 | 0.70 | Extremely Satisfied |

The statistical findings in Table 3 illustrate the high efficacy of the proposed system. Most notably, the Dashboard and Data Visibility criteria achieved a perfect mean score ($M = 5.00, \sigma = 0.00$), indicating unanimous agreement among the diverse participant group regarding its effectiveness in project oversight. Furthermore, the system was perceived as significantly more organized than traditional tracking methods ($M = 4.90, \sigma = 0.32$). These results suggest that the integration of automated task tracking and visual data analysis successfully addresses the common pain points of manual management, providing a reliable and user-friendly solution across various professional domains. The data shows that users prefer this system to traditional tracking methods (Excel, notebooks, etc.).

## 4. CONCLUSION AND RECOMMENDATIONS

This study demonstrates that the system developed through the integration of MERN Stack and React Native technologies provides a comprehensive solution to data synchronization and traceability issues encountered in modern project management processes. Usability tests and statistical analyses confirmed that the system provides a clear organizational advantage (M=4.90) compared to traditional methods and that the visual data dashboards deliver flawless performance (M=5.00, $\sigma$=0.00) in executive decision support processes. The MongoDB-based centralized architecture ensures data integrity, while cross-platform support enables workflows to continue seamlessly regardless of the office environment. Based on feedback from participants, future work will focus on the responsive design of the system's pop-up interfaces and increasing file upload capacities for high-resolution asset management. Furthermore, integrating artificial intelligence-based task prioritization algorithms into the system and developing customized reporting modules based on user roles will further strengthen the application's corporate adaptation process.

## ACKNOWLEDGEMENTS

This work is supported by the Scientific and Technological Research Council of Türkiye (TÜBİTAK) under the 2209-A University Student Research Projects Support Program (Project No: 1919B012453439, 2024/1 term). The authors are solely responsible for the content of this study, and the views expressed do not necessarily reflect the official positions of TÜBİTAK.